\documentclass[aps,prd,twocolumn,showpacs,superscriptaddress,amsmath,amssymb,nofootinbib]{revtex4-1}
\usepackage{graphicx}
\usepackage{subfigure}
\usepackage{float}
\usepackage{bigstrut}
\usepackage{overpic}
\usepackage{dcolumn}
\usepackage{bm}
\usepackage{color}
\usepackage{amsmath}
\usepackage{amsfonts}
\usepackage{amssymb}
\usepackage{lineno}
\usepackage{xspace}
\usepackage{multirow}
\usepackage{subfigure}
\usepackage{verbatim}
\usepackage{footmisc}
\usepackage{epstopdf}
\usepackage{xcolor}
\usepackage{soul}
\usepackage{comment}
\usepackage{siunitx}
\usepackage{dcolumn}

\usepackage[colorlinks,linkcolor=blue,anchorcolor=blue,citecolor=blue]{hyperref}

\hypersetup{hypertex=true,
	colorlinks=true,
	linkcolor=blue,
	anchorcolor=blue,
	citecolor=blue}

\newcommand{\mevcc}{\,\unit{MeV}/c^2}
\newcommand{\gev}{\,\unit{GeV}}
\newcommand{\gevc}{\,\unit{GeV}/c}

\newcommand{\ee}{e^{+}e^{-}}

\newcommand{\Modea}{\bar{p}{}K_{S}^0}
\newcommand{\Modeb}{\bar{p}{}K^{+}\pi^-}
\newcommand{\Modec}{\bar{p}{}K_{S}^0\pi^0}
\newcommand{\Moded}{\bar{p}{}K_{S}^0\pi^-\pi^+}
\newcommand{\Modee}{\bar{p}{}K^{+}\pi^-\pi^0}
\newcommand{\Modef}{\bar{p}{}\pi^-\pi^+}
\newcommand{\Modeaa}{\bar{\Lambda}{}\pi^-}
\newcommand{\Modebb}{\bar{\Lambda}{}\pi^-\pi^0}
\newcommand{\Modedd}{\bar{\Lambda}{}\pi^-\pi^+\pi^-}
\newcommand{\Modeaaa}{\bar{\Sigma}{}^{0}\pi^-}
\newcommand{\Modeccc}{\bar{\Sigma}{}^{-}\pi^0}
\newcommand{\Modeddd}{\bar{\Sigma}{}^{-}\pi^-\pi^+}

\newcommand{\lambdacp}{\Lambda_{c}^{+}}
\newcommand{\lambdacm}{\bar{\Lambda}{}_{c}^{-}}

\newcolumntype{d}[1]{D{.}{.}{#1}}
\newcommand{\LctoLmdKsK}{\Lambda_{c}^{+} \to \Lambda K_{S}^{0} K^{+}}
\newcommand{\LctoSigmaKsK}{\Lambda_{c}^{+} \to \Sigma^{0} K_{S}^{0} K^{+}}
\newcommand{\LctoXiKsPi}{\Lambda_{c}^{+} \to \Xi^{0} K_{S}^{0} \pi^{+}}
\newcommand{\LmdKsK}{\Lambda K_{S}^{0} K^{+}}
\newcommand{\SigmaKsK}{\Sigma^{0} K_{S}^{0} K^{+}}
\newcommand{\XiKsPi}{\Xi^{0} K_{S}^{0} \pi^{+}}\usepackage{xcolor}
\usepackage{hyperref}

\hypersetup{
    colorlinks=true,
    linkcolor=blue,
    urlcolor=blue
}
\begin{document}


\title{\boldmath  Measurement of the branching fractions of the Cabibbo-favored decays $\Lambda_{c}^{+}\to\Lambda K_{S}^{0}K^{+}$ and $\Lambda_{c}^{+}\to\Xi^{0}K_{S}^{0}\pi^{+}$ and search for $\Lambda_{c}^{+}\to\Sigma^{0} K_{S}^{0}K^{+}$ }

\author{
\begin{small}
M.~Ablikim$^{1}$, M.~N.~Achasov$^{4,c}$, P.~Adlarson$^{76}$, X.~C.~Ai$^{81}$, R.~Aliberti$^{35}$, A.~Amoroso$^{75A,75C}$, Q.~An$^{72,58,a}$, Y.~Bai$^{57}$, O.~Bakina$^{36}$, Y.~Ban$^{46,h}$
% ...
\\
\vspace{0.2cm}
(BESIII Collaboration)\\
\vspace{0.2cm} {\it
$^{1}$ Institute of High Energy Physics, Beijing 100049, People's Republic of China\\
% ...
$^{4}$ Budker Institute of Nuclear Physics SB RAS (BINP), Novosibirsk 630090, Russia\\
% ...
$^{35}$ Johannes Gutenberg University of Mainz, Johann-Joachim-Becher-Weg 45, D-55099 Mainz, Germany\\
$^{36}$ Joint Institute for Nuclear Research, 141980 Dubna, Moscow region, Russia\\
% ...
$^{46}$ Peking University, Beijing 100871, People's Republic of China\\
% ...
$^{57}$ Southeast University, Nanjing 211100, People's Republic of China\\
$^{58}$ State Key Laboratory of Particle Detection and Electronics, Beijing 100049, Hefei 230026, People's Republic of China\\
% ...
$^{72}$ University of Science and Technology of China, Hefei 230026, People's Republic of China\\
% ...
$^{75}$ University of Turin and INFN, (A)University of Turin, I-10125, Turin, Italy; (B)University of Eastern Piedmont, I-15121, Alessandria, Italy; (C)INFN, I-10125, Turin, Italy\\
$^{76}$ Uppsala University, Box 516, SE-75120 Uppsala, Sweden\\
% ...
$^{81}$ Zhengzhou University, Zhengzhou 450001, People's Republic of China\\
\vspace{0.2cm}
$^{a}$ Deceased\\
% ...
$^{c}$ Also at the Novosibirsk State University, Novosibirsk, 630090, Russia\\
% ...
$^{h}$ Also at State Key Laboratory of Nuclear Physics and Technology, Peking University, Beijing 100871, People's Republic of China\\
% ...
}
\vspace{0.4cm}
\end{small}
}

\vspace{4cm}

\date{\it \small \bf \today}

\begin{abstract}
Based on $e^{+}e^{-}$ collision data corresponding to an integrated luminosity of about 4.5\,fb$^{-1}$ collected at center-of-mass energies between $4599.53$~MeV and $4698.82$~MeV with the BESIII detector, the absolute branching fraction of the Cabibbo-favored decay $\Lambda_{c}^{+}\to\Lambda K_{S}^{0}K^{+}$ is measured to be $(3.12\pm0.46\pm0.15)\times10^{-3}$. Combined with a previous measurement from the BESIII Collaboration, the branching fraction of the decay $\Lambda_{c}^{+}\to\Lambda K_{S}^{0}K^{+}$ is calculated to be $(3.07\pm0.26\pm0.13)\times10^{-3}$. The decay $\Lambda_{c}^{+}\to\Xi^{0}K_{S}^{0}\pi^{+}$ is observed for the first time with a statistical significance of $6.6\sigma$, and its branching fraction is determined to be 
$(3.70\pm0.60\pm0.21)\times10^{-3}$. In addition, a search for the decay  $\Lambda_{c}^{+}\to\Sigma^{0} K_{S}^{0}K^{+}$ is performed and its branching fraction is determined to be $(0.80^{+0.28}_{-0.24}\pm0.16)\times10^{-3}$, corresponding to an upper limit of $1.28\times10^{-3}$ at $90\%$ confidence level.
These measurements provide new information that can be used to distinguish between theoretical models.

\end{abstract}

\maketitle


\section{introduction}
Hadronic decays of charmed baryons serve as an ideal laboratory to understand the weak and strong interactions in the charm sector~\cite{ref:Lu2016ogy}. The branching fractions (BFs) of charmed baryon decays can provide essential input for constraining theoretical models~\cite{ref:Geng2018rse}. 
Decay amplitudes of charmed baryons can be separated into factorizable terms, allowing for the independent treatment of the strong and weak components, and non-factorizable terms, where it is difficult to calculate the interplay between the two components~\cite{Chau:1982da, Chau:1995g}.
The external \emph{W}-emission diagram is predominantly factorizable, while internal \emph{W}-emission and \emph{W}-exchange diagrams incorporate non-factorizable terms, which contribute to the heightened complexity of theoretical calculations for charmed baryon decays.

\begin{table*}[t]
	\caption{The theoretical predictions and experimental results for the BFs of $\LctoLmdKsK$, $\LctoSigmaKsK$ and $\LctoXiKsPi$.} 
	\small
	\centering
	\begin{tabular}{l|c|c|c}
		\hline \hline
		&$\LctoLmdKsK$&$\LctoSigmaKsK$&$\LctoXiKsPi$  \\ \hline
		Geng~\cite{Geng:2018upx}&$(2.85\pm0.55)\times10^{-3}$&$(2.50\pm0.50)\times10^{-3}$&$(4.35\pm0.85)\times10^{-2}$\\ \hline
		Cen~\cite{Cen:2019ims}&$(2.90\pm0.50)\times10^{-3}$&$(0.14\pm0.05)\times10^{-3}$&$(2.20\pm0.40)\times10^{-2}$\\ \hline
		Geng~\cite{Geng:2024h5}&$(2.73\pm0.49)\times10^{-3}$&$(1.27\pm0.48)\times10^{-4}$&$(4.85\pm1.65)\times10^{-3}$\\ \hline
  
Statistical isospin model~\cite{ref1} &-&-&$(3.10\pm0.30)\times10^{-3}$\\ \hline
  
  PDG value~\cite{ref:pdg2023} &$(2.80\pm0.55)\times10^{-3}$&-&-\\ \hline
		This work &$(3.12\pm0.46\pm0.15)\times10^{-3}$&$(0.80^{+0.28}_{-0.24}\pm0.16)\times10^{-3}$&$(3.70\pm0.60\pm0.21)\times10^{-3}$ \\ \hline \hline
	\end{tabular}
	\label{tab:Th_bf_vs_exp}
\end{table*}

Numerous theoretical models and approaches have been developed to study charmed baryon decays. One possibility, a model-independent approach based on SU$(3)$-flavor symmetry, has been proposed to describe these decays, incorporating non-factorizable effects~\cite{Geng:2018upx}. While SU$(3)$ symmetry is only approximate, it is a powerful tool for extracting valuable information from the quark transitions in charmed baryon decays. Improved measurements of charmed baryon decays will be essential for further testing the SU$(3)$ symmetry assumption.
 
The ground state of the charmed baryon $\Lambda_{c}^{+}$ was first observed at the Mark~II experiment in 1979~\cite{ref:Abrams1979iu}.
Since 2014, BESIII has accumulated a large data sample at the $\Lambda_{c}^{+}\bar{\Lambda}_{c}^{-}$ threshold.
Based on these data, the BESIII experiment reported a series of BFs of exclusive decays of the $\Lambda_{c}^{+}$ baryon~\cite{BESIII:lmdenv, BESIII:hadron, BESIII:phh, BESIII:ppi0peta, BESIII:lmdmunv, BESIII:sigmapipi, BESIII:nkspi, BESIII:xik, BESIII:sigmaeta, BESIII:Lmdpieta, BESIII:pkseta, BESIII:npi}. 
Several new decay modes were discovered, and the precision of BFs for the known decay modes has also been significantly improved.
However, decay processes involving two or three strange hadrons in the final state have rarely been studied. A comparison between the BFs of inclusive and summed exclusive decay channels of $\Lambda_{c}^{+}\to\Lambda X$~\cite{ref:7, ref:pdg2023} and $\Lambda_{c}^{+}\to K_{S}^{0} X$~\cite{ref:8}, where $X$ means all possible final state particles, shows that there is still room for unmeasured decay channels involving two or three strange hadrons.
 
\begin{figure*}[!hpbt]
		\centering
		\subfigure[\ Internal \emph{W}-emission]
		{
			\includegraphics[width=0.3\textwidth]{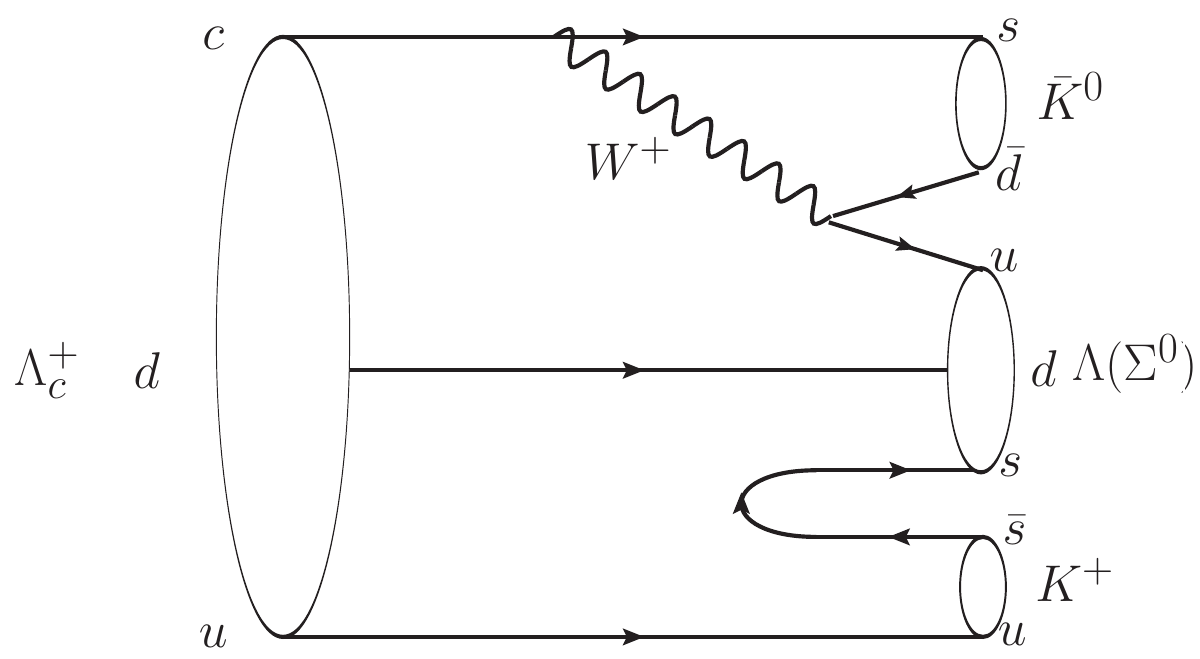}
		}
		\subfigure[\ \emph{W}-exchange]
		{
			\includegraphics[width=0.3\textwidth]{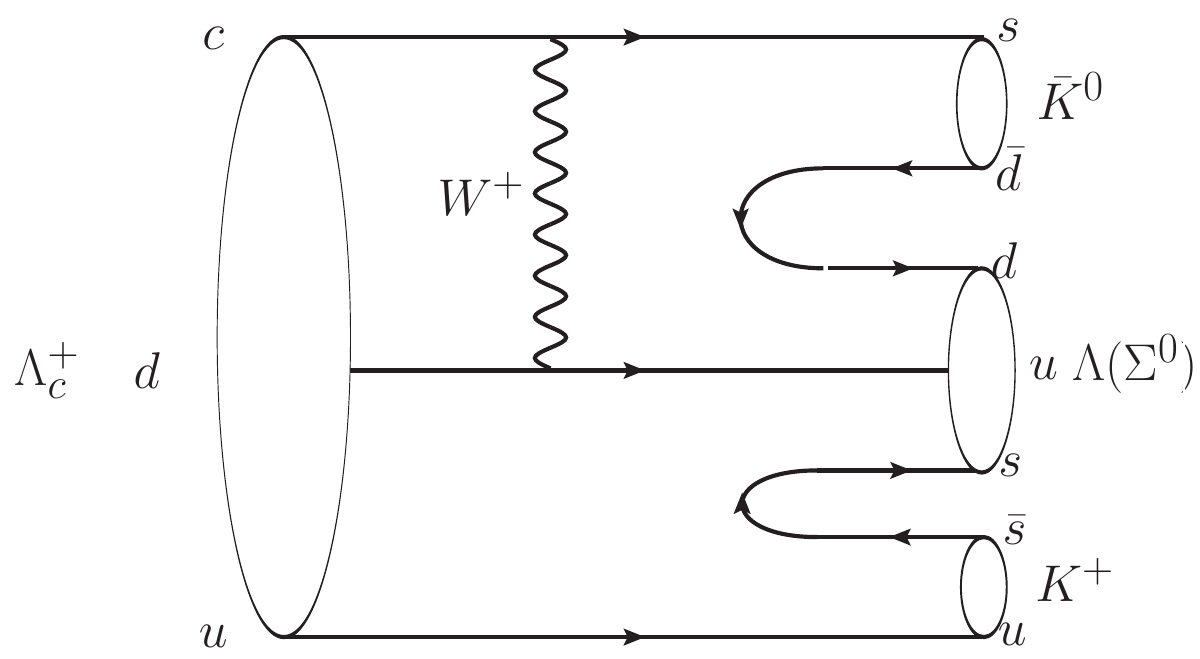}
		}
		\caption{Topological diagrams of $\Lambda_{c}^{+}\to\Lambda(\Sigma^{0})K_{S}^{0}K^{+}$.}
		\label{fig:feynman_lmdksk}
\end{figure*}

\begin{figure*}[!hpbt]
		\centering
		\subfigure[\ External \emph{W}-emission]
		{
			\includegraphics[width=0.3\textwidth]{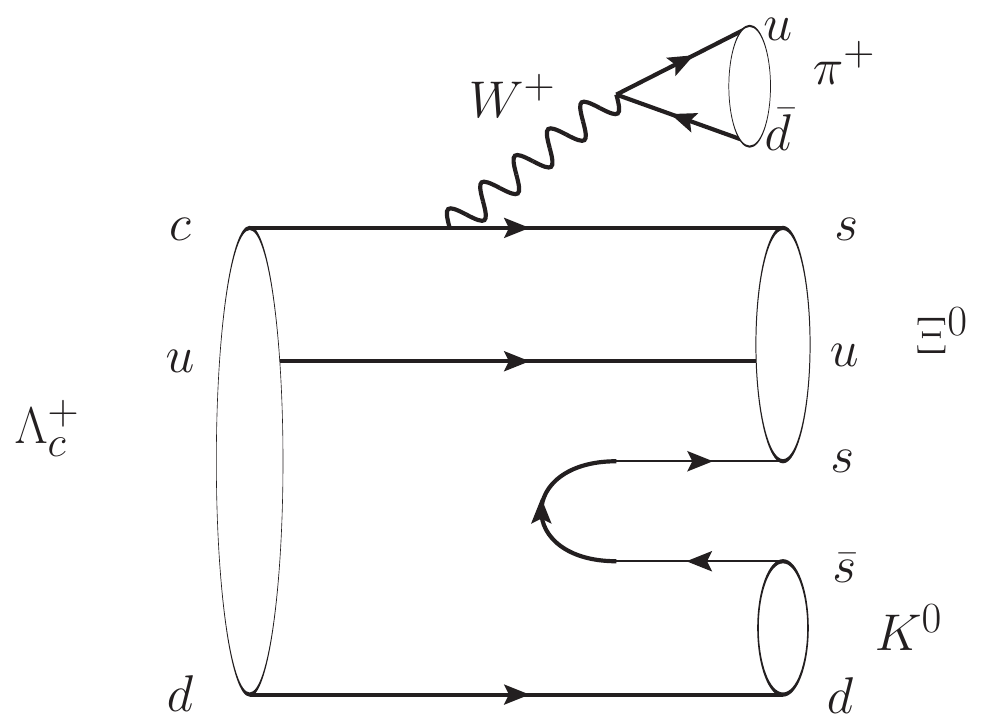}
		}
		\subfigure[\ Internal \emph{W}-emission]
		{
			\includegraphics[width=0.3\textwidth]{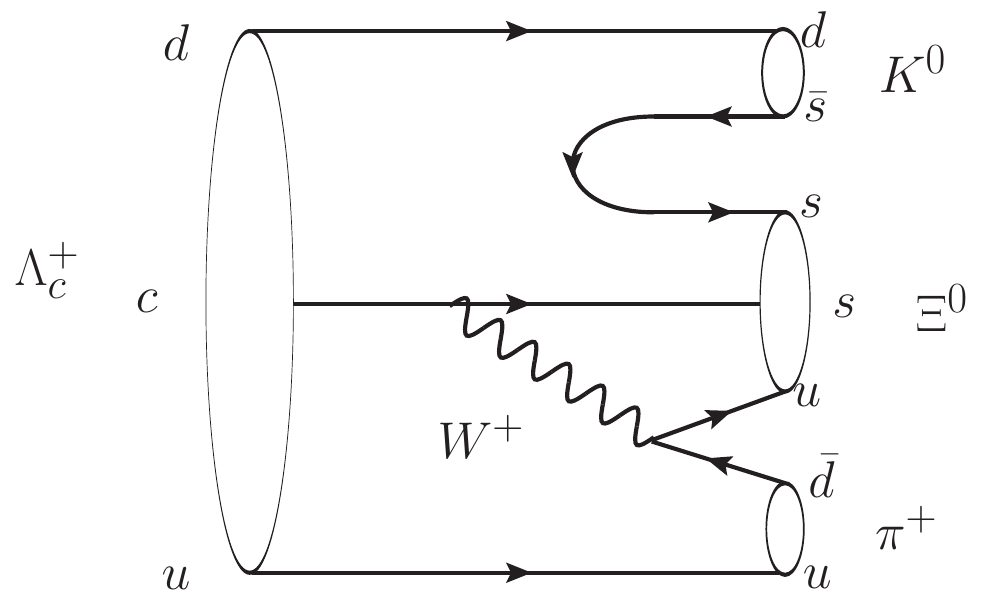}
		}
		\subfigure[\ \emph{W}-exchange]
		{
			\includegraphics[width=0.3\textwidth]{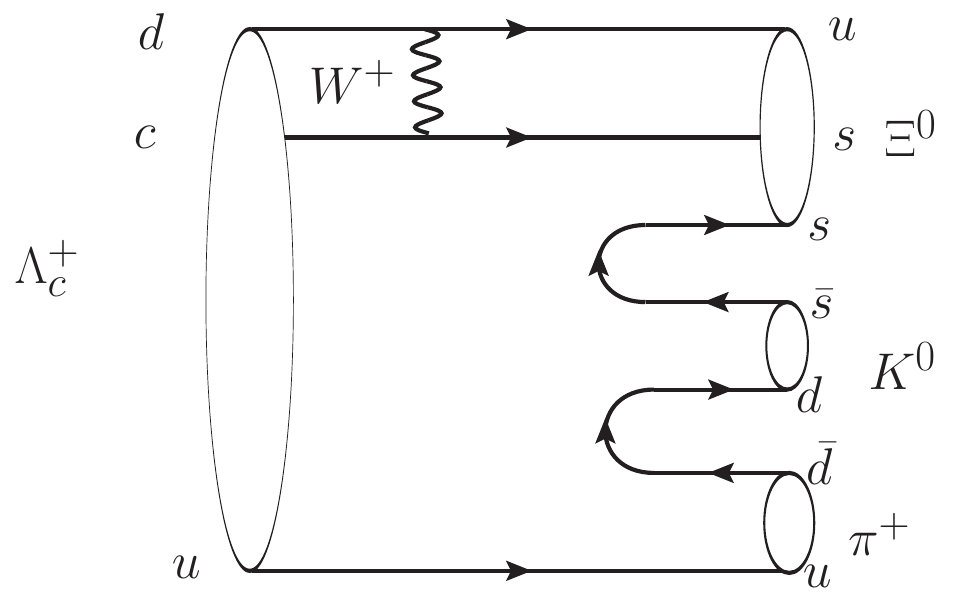}
		}
		\caption{Topological diagrams of $\Lambda_{c}^{+}\to \Xi^{0} K_{S}^{0} \pi^{+}$.}
		\label{fig:feyaman_xikspi}
	\end{figure*}

 The decays $\LctoLmdKsK$, $\LctoSigmaKsK$ and $\LctoXiKsPi$ are predominantly characterized by the $c \to s$ transition but involve more than one strange hadron. The topology diagrams illustrating the three decay modes are displayed in Figs.~\ref{fig:feynman_lmdksk} and \ref{fig:feyaman_xikspi}. Predictions for the BFs of these decay modes by SU(3)-flavor symmetry or by the statistical isospin model are listed in Table~\ref{tab:Th_bf_vs_exp} alongside the averaged results from the Particle Data Group~(PDG)~\cite{ref:pdg2023}.
 Further measurements of the BFs of these decays are needed to help confirm the predictions of SU(3)-flavor symmetry and to offer valuable information for further theoretical studies~\cite{Geng:2018upx}.

%

In this analysis, we conduct precise measurements of the absolute BFs of $\LctoLmdKsK$ and $\LctoXiKsPi$, and search for $\LctoSigmaKsK$  by  employing  the  double-tag  (DT)  method~\cite{mark3}. Our analysis uses electron-positron collision data samples collected at seven center-of-mass energies ($E_{\rm cm}$) ranging from $4599.53$~MeV to $4698.82$~MeV by the BESIII detector~\cite{BESIII:2022ulv}.
These data samples correspond to an integrated luminosity of 4.5\,fb$^{-1}$~\cite{BESIII:2022ulv}, as detailed in Table~\ref{tab:luminosity}.  
Throughout this paper, charge conjugation is always implied.
\begin{table}[h]
  \begin{center}
    \caption{Center-of-mass energies and luminosities of the data samples used in this work.}
    \sisetup{uncertainty-mode=separate}
    \begin{tabular}{S[table-format=5.5(3)]|S[table-format=4.3(1)]}
      \hline \hline
      {$E_{\rm cm}$ (MeV)}	&  {Luminosity (pb$^{-1}$)} \\ \hline
      4599.53\pm0.07\pm0.74   &   586.9\pm0.1\pm3.9  \\
      4611.86\pm0.12\pm0.32   &   103.8\pm0.1\pm0.6  \\
      4628.00\pm0.06\pm0.32   &   521.5\pm0.1\pm2.8  \\
      4640.91\pm0.06\pm0.38   &   552.4\pm0.1\pm2.9  \\
      4661.24\pm0.06\pm0.29   &   529.6\pm0.1\pm2.8  \\
      4681.92\pm0.08\pm0.29   &   1669.3\pm0.2\pm8.8 \\
      4698.82\pm0.10\pm0.39   &   536.4\pm0.1\pm2.8  \\
      \hline \hline
    \end{tabular}
    \label{tab:luminosity}
  \end{center}
\end{table}

\section{BESIII DETECTOR AND MONTE CARLO SIMULATION}
The BESIII detector~\cite{BESIII:2009fln} records symmetric $\ee$ collisions provided by the BEPCII storage ring~\cite{ref:14}, in the center-of-mass energy ranging from $1.85$ to $4.95$~GeV, with a peak luminosity of $1.1\times10^{33}\mathrm{cm}^{-2}\mathrm{s}^{-1}$ achieved in 2023 at a center-of-mass energy of $\sqrt{s}=3.773$~GeV. BESIII has collected large data samples in this energy region~\cite{BESIII:2020nme}. The cylindrical core
 of the BESIII detector covers $93\%$ of the full solid angle
 and is comprised of a helium-based multilayer drift chamber
 (MDC), a plastic scintillator time-of-flight system (TOF),
 and a CsI(Tl) electromagnetic calorimeter (EMC), which
 are all enclosed in a superconducting solenoidal magnet
 providing a 1.0~T magnetic field.
 The solenoid is supported by an octagonal flux-return yoke
which is segmented into layers and instrumented with
resistive plate counter modules for muon identification.
  The charged-particle momentum resolution at 1~$\gevc$
 is $0.5\%$, and the specific ionization energy loss
 ($\mathrm{d}E/\mathrm{d}x$) resolution is $6\%$ for electrons from Bhabha scattering. The EMC measures photon energies with a resolution of $2.5\%(5\%)$ at 1~$\gev$ in the barrel (end-cap) region. The time resolution of the TOF in the barrel region is 68~ps, while that in the end-cap region is 110~ps.
 The end-cap TOF system was upgraded in 2015 using multi-gap resistive plate chamber technology, providing a time resolution of 60~ps~\cite{Li:2017jpg, Guo:2017sjt, Cao:2020ibk}. About $85\%$ of the $\Lambda_{c}^{+}\bar{\Lambda}_{c}^{-}$ pairs are produced in data taken after this upgrade. More detailed descriptions can be found in Refs.~\cite{BESIII:2009fln, ref:14}.

 High-statistics Monte Carlo (MC) simulation samples for the annihilation of $\ee$ are produced with the {\footnotesize{KKMC}} generator~\cite{ref:Jadach2000ir} by incorporating the initial-state radiation (ISR) effects and the beam-energy spread. The inclusive MC sample, which consists of $\lambdacp\lambdacm$ events, $D^{(*)}_{(s)}$ production, ISR return to the lower-mass $\psi$ states, and the continuum processes $\ee\to q \bar{q}$ $(q=u,d,s)$, is generated to determine the single-tag (ST) detection efficiency and estimate the potential background.
All particle decays are modeled with {\sc evtgen}~\cite{ref:Lange2001uf, ref:Ping2008zz} using BFs either taken from the
PDG~\cite{ref:pdg2023}, when available, or otherwise estimated with {\sc lundcharm}~\cite{Chen:2000tv}.
 Furthermore, exclusive DT signal MC events, where the $\lambdacm$ decays into any of the tag modes and the $\lambdacp$ decays into any of the signal modes are used to determine the DT detection efficiencies. The Born cross sections are taken into account when producing the MC sample of $\lambdacp\lambdacm$ pairs~\cite{BESIII:2023rwv}.
 The simulated $\LctoLmdKsK$ and $\LctoSigmaKsK$ signal MC samples are modeled uniformly distributed in phase space (PHSP), and the resulting angular, momentum and two-body invariant mass distributions of the final state particles are in a good agreement with the data.
 For the signal simulated sample of $\LctoXiKsPi$, the key kinematic distributions mentioned above have been reweighted to agree with those of data.
 All final tracks are fed into a {\footnotesize{GEANT4}} based detector simulation package, which includes the geometric description of the BESIII detector and also generates showers~\cite{GEANT4:2002zbu, Allison:2006ve}.

\section{event selection}

We first reconstruct a $\lambdacm$ baryon candidate, referred to as the ST candidate, using one of the twelve exclusive decay modes listed in Table~\ref{tab:STyields}.
Then, we reconstruct the signal decays $\LctoLmdKsK$, $\LctoSigmaKsK$ and $\LctoXiKsPi$, referred to as the DT candidates, in the system recoiling against the $\bar{\Lambda}_{c}^{-}$ ST candidate. The DT events are selected using a partial reconstruction technique, as described below. The selection criteria for the $\lambdacm$ ST candidates of the twelve tag modes are the same as those described in Ref.~\cite{BESIII:2022xne}.

For $\LctoLmdKsK$ and $\LctoSigmaKsK$, we search for a candidate $K_{S}^{0}$ and $K^{+}$, while for $\LctoXiKsPi$, a candidate $K_{S}^{0}$ and $\pi^{+}$ are reconstructed.
The $K_{S}^{0}$ particle candidate is formed by combining two oppositely charged tracks, which satisfy $|\mathrm{cos\theta|<0.93}$ (where the angle $\theta$ is defined with respect to the $z$-axis, which is the symmetry axis of the MDC) and their distances of closest approach to the interaction point~(IP) are required to be within $\pm20$~cm along the beam direction. No additional requirements are made for particle identification~(PID) and distance in the transverse plane for the two tracks.
 To veto backgrounds from the mis-combination of pions, the two daughter tracks from the $K_{S}^{0}$ are required to originate from a common decay vertex by applying a vertex fit and the $\chi^{2}$ of the vertex fit must be less than 100.
Furthermore, since the $K_{S}^{0}$ has a relatively long life time, the decay vertex is required to be separated from the IP by a distance of at least twice the corresponding uncertainty ($\sigma_{L}$) of the decay length ($L$).
Tracks satisfying the above requirements are assigned as the $\pi^{+}$ and $\pi^{-}$ candidates from the $K_{S}^{0}$ decay.
The fitted momenta of the $\pi^{+}\pi^{-}$ combinations are  utilized in the subsequent analysis.
The invariant mass of the $\pi^+\pi^-$ combination $M(\pi^{+}\pi^{-})$ must lie in the mass range $487\mevcc<M(\pi^{+}\pi^{-})<511\mevcc$, corresponding to a window of about three times the resolution.
If multiple $K_{S}^{0}$ particle candidates are reconstructed in the event, the one with the maximum $L/\sigma_{L}$ is chosen.
We have neglected the exchange of the two $K_{S}^{0}$ candidates in the $\lambdacp\lambdacm$ events since it is found to be negligible based on the inclusive MC sample, and the background contribution is represented by a flat distribution.

An additional charged particle is selected, with the requirements that the track must satisfy $|\mathrm{cos\theta}|<0.93$ and have a distance  of closest approach to the IP less than 10~cm along the beam axis and less than 1~cm in the perpendicular plane.
This particle is identified as a kaon or a pion candidate by comparing the likelihoods $\mathcal{L}$ with the different hypotheses calculated by the energy loss in the MDC and the flight time measured by the TOF system, where the requirement for $K$ is $\mathcal{L}(K)>\mathcal{L}(\pi)$ and for $\pi$ is $\mathcal{L}(\pi)>\mathcal{L}(K)$.

\section{data analysis}

\sisetup{uncertainty-mode=separate}
 \begin{table*}[t]
  \begin{center}
   \caption{ST yield ($N_{i}^{\rm ST}$) of 12~modes from $4599.53$ to $4698.82$~MeV. The uncertainty is statistical only.}
  \footnotesize
  \begin{tabular}{l|S[table-format=4(2)]|
    S[table-format=4(2)]|S[table-format=4(2)]|S[table-format=4(2)]|S[table-format=4(2)]|S[table-format=5(3)]|S[table-format=4(2)]}
      \hline \hline
  Mode  & {$4599.53$\,MeV} & {$4611.86$\,MeV}  & {$4628.00$\,MeV}  & {$4640.91$\,MeV} & {$4661.24$\,MeV} & {$4681.92$\,MeV} & {$4698.82$\,MeV} \\ \hline
$\textbf{$\Modea$}$   & 1277\pm36 & 239\pm16  & 1062\pm35 & 1116\pm36 & 1119\pm35 & 3376\pm61   & 958\pm33 \\ 
$\textbf{$\Modeb$}$   & 6806\pm91 & 1166\pm39 & 5969\pm89 & 6337\pm90 & 5938\pm86 & 17508\pm147 & 5167\pm80 \\ 
$\textbf{$\Modec$}$   & 606\pm34  & 127\pm17  & 624\pm36  & 609\pm36  & 594\pm36  & 1785\pm63   & 471\pm34 \\ 
$\textbf{$\Moded$}$   & 613\pm34  & 106\pm16  & 517\pm33  & 560\pm35  & 537\pm33  & 1511\pm57   & 462\pm31 \\ 
$\textbf{$\Modee$}$   & 2197\pm78 & 364\pm34  & 1615\pm69 & 1662\pm72 & 1700\pm73 & 5111\pm128  & 1389\pm74 \\ 
$\textbf{$\Modeaa$}$  & 757\pm28  & 123\pm11  & 682\pm29  & 712\pm29  & 668\pm27  & 2074\pm48   & 538\pm25 \\ 
$\textbf{$\Modebb$}$  & 1743\pm56 & 302\pm23  & 1474\pm54 & 1639\pm55 & 1491\pm51 & 4380\pm88   & 1301\pm49 \\ 
$\textbf{$\Modedd$}$  & 768\pm36  & 139\pm15  & 595\pm33  & 756\pm37  & 780\pm36  & 2059\pm61   & 639\pm34 \\ 
$\textbf{$\Modeaaa$}$ & 520\pm26  & 102\pm13  & 419\pm24  & 452\pm25  & 454\pm25  & 1398\pm42   & 371\pm22 \\ 
$\textbf{$\Modeccc$}$ & 320\pm25  & 73\pm10   & 267\pm23  & 303\pm25  & 298\pm25  & 879\pm43    & 251\pm24 \\ 
$\textbf{$\Modeddd$}$ & 1186\pm49 & 218\pm22  & 1094\pm49 & 1094\pm50 & 1066\pm49 & 3027\pm88   & 956\pm48 \\ 
$\textbf{$\Modef$}$   & 598\pm47  & 155\pm22  & 524\pm45  & 559\pm48  & 590\pm48  & 1596\pm80   & 459\pm47 \\
\hline \hline
   \end{tabular}
   \label{tab:STyields}
  \end{center}
  \end{table*}
\begin{table*}[t]
  \begin{center}
   \caption{ST detection efficiency ($\varepsilon_{i}^{\rm ST}$, in $\%$) of 12~modes from $4599.53$ to $4698.82$~MeV. The uncertainty is statistical only.}
  \footnotesize
  \begin{tabular}{l|c|c|c|c|c|c|c}
      \hline \hline
  Mode  & $4599.53$\,MeV & $4611.86$\,MeV  & $4628.00$\,MeV  & $4640.91$\,MeV & $4661.24$\,MeV & $4681.92$\,MeV & $4698.82$\,MeV \\ \hline
  $\textbf{$\Modea$}$   & $56.1\pm0.2$ & $53.7\pm0.5$ & $52.2\pm0.2$ & $51.1\pm0.2$ & $49.6\pm0.2$ & $48.6\pm0.1$ & $47.5\pm0.2$ \\ 
$\textbf{$\Modeb$}$   & $51.5\pm0.1$ & $51.0\pm0.2$ & $49.9\pm0.1$ & $49.2\pm0.1$ & $48.2\pm0.1$ & $47.1\pm0.1$ & $46.3\pm0.1$ \\ 
$\textbf{$\Modec$}$   & $23.0\pm0.2$ & $22.2\pm0.4$ & $21.0\pm0.2$ & $21.0\pm0.2$ & $20.1\pm0.2$ & $19.7\pm0.1$ & $18.9\pm0.2$ \\ 
$\textbf{$\Moded$}$   & $23.5\pm0.2$ & $21.9\pm0.5$ & $20.9\pm0.2$ & $21.1\pm0.2$ & $20.2\pm0.2$ & $20.7\pm0.1$ & $19.5\pm0.2$ \\ 
$\textbf{$\Modee$}$   & $20.6\pm0.1$ & $19.9\pm0.2$ & $19.0\pm0.1$ & $18.4\pm0.1$ & $18.1\pm0.1$ & $17.8\pm0.1$ & $17.5\pm0.1$ \\ 
$\textbf{$\Modeaa$}$  & $48.4\pm0.3$ & $46.9\pm0.6$ & $43.7\pm0.3$ & $43.1\pm0.3$ & $41.7\pm0.3$ & $41.3\pm0.2$ & $39.4\pm0.3$ \\ 
$\textbf{$\Modebb$}$  & $21.6\pm0.1$ & $19.8\pm0.2$ & $19.4\pm0.1$ & $19.4\pm0.1$ & $18.9\pm0.1$ & $18.2\pm0.1$ & $17.9\pm0.1$ \\ 
$\textbf{$\Modedd$}$  & $15.6\pm0.1$ & $13.6\pm0.3$ & $14.1\pm0.1$ & $14.4\pm0.1$ & $14.1\pm0.1$ & $13.9\pm0.1$ & $14.6\pm0.1$ \\ 
$\textbf{$\Modeaaa$}$ & $29.4\pm0.2$ & $26.6\pm0.6$ & $27.6\pm0.2$ & $26.6\pm0.2$ & $26.3\pm0.2$ & $25.9\pm0.1$ & $24.3\pm0.2$ \\ 
$\textbf{$\Modeccc$}$ & $23.7\pm0.3$ & $22.6\pm0.6$ & $23.8\pm0.3$ & $25.0\pm0.3$ & $23.2\pm0.3$ & $22.5\pm0.2$ & $22.2\pm0.3$ \\ 
$\textbf{$\Modeddd$}$ & $25.4\pm0.1$ & $25.5\pm0.3$ & $24.0\pm0.1$ & $23.7\pm0.1$ & $23.2\pm0.1$ & $22.1\pm0.1$ & $22.2\pm0.1$ \\ 
$\textbf{$\Modef$}$   & $64.3\pm0.3$ & $62.7\pm0.8$ & $62.4\pm0.4$ & $60.4\pm0.4$ & $60.2\pm0.4$ & $53.6\pm0.2$ & $55.9\pm0.4$ \\
\hline \hline
   \end{tabular}
   \label{tab:STeff}
  \end{center}
  \end{table*}
\begin{table*}[t]
	\begin{center}
		\caption{DT efficiencies  of $\Lambda_{c}^{+}\to\Lambda K_{S}^{0}K^{+}$  ($\varepsilon_{i,\Lambda K_{S}^{0}K^{+}}^{\rm DT}$, in $\%$)
  for 12~tag modes at various energy points. The uncertainty is statistical only.}
		  \footnotesize
		\begin{tabular}{l|S[table-format=2.2(2)]|
    S[table-format=2.2(2)]|S[table-format=2.2(2)]|S[table-format=2.2(2)]|S[table-format=2.2(2)]|S[table-format=2.2(2)]|S[table-format=2.2(2)]}
			\hline \hline
  Mode & {4599.53 MeV}       & {4611.86 MeV}       & {4628.00 MeV}       & {4640.91 MeV}       & {4661.24 MeV}       & {4681.92 MeV}       & {4698.82 MeV}       \\ \hline
			$\textbf{$\Modea$}$   & 15.01\pm0.18 & 13.68\pm0.17 & 13.33\pm0.17 & 13.21\pm0.17 & 12.83\pm0.17 & 12.86\pm0.17 & 12.34\pm0.17 \\ 
			$\textbf{$\Modeb$}$   & 13.98\pm0.12 & 12.79\pm0.12 & 12.41\pm0.12 & 12.61\pm0.12 & 12.61\pm0.12 & 12.20\pm0.12 & 12.00\pm0.12 \\ 
			$\textbf{$\Modec$}$   & 5.89\pm0.11 & 5.23\pm0.10 & 5.23\pm0.10 & 5.39\pm0.10 & 5.27\pm0.10 & 4.96\pm0.10 & 5.09\pm0.10 \\ 
			$\textbf{$\Moded$}$   & 5.42\pm0.11 & 4.59\pm0.11 & 4.38\pm0.10 & 4.52\pm0.10 & 4.49\pm0.10 & 4.39\pm0.10 & 4.40\pm0.10 \\ 
			$\textbf{$\Modee$}$   & 5.37\pm0.10 & 4.82\pm0.09 & 5.02\pm0.09 & 4.99\pm0.09 & 4.76\pm0.09 & 4.42\pm0.09 & 4.50\pm0.09 \\ 
			$\textbf{$\Modeaa$}$  & 12.45\pm0.26 & 11.28\pm0.25 & 10.80\pm0.25 & 10.92\pm0.25 & 10.73\pm0.24 & 10.45\pm0.24 & 10.56\pm0.24 \\ 
			$\textbf{$\Modebb$}$  & 5.31\pm0.08 & 4.63\pm0.07 & 4.62\pm0.07 & 4.52\pm0.07 & 4.57\pm0.07 & 4.37\pm0.07 & 4.19\pm0.07 \\ 
			$\textbf{$\Modedd$}$  & 3.43\pm0.09 & 2.92\pm0.08 & 3.17\pm0.08 & 3.02\pm0.08 & 2.97\pm0.08 & 3.12\pm0.08 & 2.97\pm0.08 \\ 
			$\textbf{$\Modeaaa$}$ & 7.75\pm0.21 & 7.58\pm0.21 & 6.72\pm0.20 & 7.00\pm0.20 & 6.73\pm0.20 & 6.65\pm0.20 & 6.35\pm0.19 \\ 
			$\textbf{$\Modeccc$}$ & 6.42\pm0.20 & 6.52\pm0.20 & 6.29\pm0.20 & 5.71\pm0.19 & 5.99\pm0.19 & 5.61\pm0.18 & 5.76\pm0.19 \\ 
			$\textbf{$\Modeddd$}$ & 6.90\pm0.11 & 6.30\pm0.10 & 6.20\pm0.10 & 6.18\pm0.10 & 6.04\pm0.10 & 5.97\pm0.10 & 5.54\pm0.10 \\ 
			$\textbf{$\Modef$}$   & 15.50\pm0.48 & 14.59\pm0.47 & 14.45\pm0.47 & 14.14\pm0.46 & 13.83\pm0.45 & 13.14\pm0.45 & 13.22\pm0.45 \\ \hline \hline
		\end{tabular}
		
		\label{tab:DTeff_lmdksk}
	\end{center}
\end{table*}

\begin{table*}[t]
  \begin{center}
          \caption{DT efficiencies of $\Lambda_{c}^{+}\to\Sigma^{0} K_{S}^{0}K^{+}$ ($\varepsilon_{i,\Sigma^{0} K_{S}^{0}K^{+}}^{\rm DT}$, in $\%$)
            for 12~tag modes at various energy points. The uncertainty is statistical only.}
          \footnotesize
          \sisetup{uncertainty-mode=separate}
          
          \begin{tabular}{l|S[table-format=2.2(2)]|
    S[table-format=2.2(2)]|S[table-format=2.2(2)]|S[table-format=2.2(2)]|S[table-format=2.2(2)]|S[table-format=2.2(2)]|S[table-format=2.2(2)]}
            \hline \hline
            Mode & {4599.53 MeV}      & {4611.86 MeV}       & {4628.00 MeV}       & {4640.91 MeV}       & {4661.24 MeV}       & {4681.92 MeV}       & {4698.82 MeV}       \\ \hline
            $\textbf{$\Modea$}$   & 10.97\pm0.16 & 10.08\pm0.15 & 9.89\pm0.15 & 9.93\pm0.15 & 9.69\pm0.15 & 9.75\pm0.15 & 9.52\pm0.15 \\ 
            $\textbf{$\Modeb$}$   & 10.31\pm0.11 & 9.30\pm0.10 & 9.40\pm0.10 & 9.39\pm0.10 & 9.20\pm0.10 & 9.21\pm0.10 & 9.18\pm0.10 \\ 
            $\textbf{$\Modec$}$   & 4.45\pm0.09 & 3.94\pm0.09 & 3.94\pm0.09 & 3.94\pm0.09 & 3.92\pm0.09 & 3.87\pm0.09 & 3.90\pm0.09 \\ 
            $\textbf{$\Moded$}$   & 3.89\pm0.10 & 3.18\pm0.09 & 3.04\pm0.09 & 3.18\pm0.09 & 3.17\pm0.09 & 3.27\pm0.09 & 3.07\pm0.09 \\ 
            $\textbf{$\Modee$}$   & 4.13\pm0.09 & 3.69\pm0.08 & 3.71\pm0.08 & 3.68\pm0.08 & 3.74\pm0.08 & 3.56\pm0.08 & 3.46\pm0.08 \\ 
            $\textbf{$\Modeaa$}$  & 9.52\pm0.23 & 8.20\pm0.22 & 8.06\pm0.21 & 8.04\pm0.21 & 7.54\pm0.21 & 7.78\pm0.21 & 7.90\pm0.21 \\ 
            $\textbf{$\Modebb$}$  & 3.85\pm0.06 & 3.49\pm0.06 & 3.46\pm0.06 & 3.52\pm0.06 & 3.38\pm0.06 & 3.37\pm0.06 & 3.37\pm0.06 \\ 
            $\textbf{$\Modedd$}$  & 2.56\pm0.07 & 2.12\pm0.07 & 2.04\pm0.07 & 2.26\pm0.07 & 2.44\pm0.07 & 2.25\pm0.07 & 2.35\pm0.07 \\ 
            $\textbf{$\Modeaaa$}$ & 6.16\pm0.19 & 5.04\pm0.17 & 5.01\pm0.17 & 5.01\pm0.17 & 5.23\pm0.18 & 5.02\pm0.17 & 4.70\pm0.17 \\ 
            $\textbf{$\Modeccc$}$ & 5.10\pm0.18 & 4.84\pm0.17 & 4.86\pm0.17 & 4.51\pm0.17 & 4.61\pm0.17 & 4.13\pm0.16 & 4.45\pm0.17 \\ 
            $\textbf{$\Modeddd$}$ & 5.07\pm0.09 & 4.73\pm0.09 & 4.67\pm0.09 & 4.57\pm0.09 & 4.47\pm0.09 & 4.56\pm0.09 & 4.49\pm0.09 \\ 
            $\textbf{$\Modef$}$   & 11.08\pm0.42 & 10.78\pm0.41 & 11.14\pm0.42 & 11.36\pm0.43 & 9.93\pm0.40 & 11.09\pm0.42 & 9.92\pm0.39 \\ \hline \hline
          \end{tabular}
          
          \label{tab:DTeff_sigmaksk}
	\end{center}
      \end{table*}
      \begin{table*}[t]
	\begin{center}
          \caption{DT efficiencies of $\Lambda_{c}^{+}\to \Xi^{0}K_{S}^{0}\pi^{+}$ ($\varepsilon_{i,\Xi^{0}K_{S}^{0}\pi^{+}}^{\rm DT}$, in $\%$)
            for 12~tag modes at various energy points.  The uncertainty is statistical only.}
          \footnotesize
          
          \begin{tabular}{l|S[table-format=2.2(2)]|
            S[table-format=2.2(2)]|S[table-format=2.2(2)]|S[table-format=2.2(2)]|S[table-format=2.2(2)]|S[table-format=2.2(2)]|S[table-format=2.2(2)]}
            \hline \hline
            Mode	& {4599.53 MeV}      & {4611.86 MeV}       & {4628.00 MeV}       & {4640.91 MeV}       & {4661.24 MeV}       & {4681.92 MeV}       & {4698.82 MeV}     \\ \hline
            $\textbf{$\Modea$}$   & 21.98\pm0.21 & 19.03\pm0.20 & 19.95\pm0.20 & 20.17\pm0.20 & 20.26\pm0.20 & 20.95\pm0.21 & 20.62\pm0.20 \\ 
            $\textbf{$\Modeb$}$   & 19.65\pm0.14 & 18.18\pm0.14 & 18.53\pm0.14 & 18.58\pm0.14 & 18.62\pm0.14 & 18.87\pm0.14 & 18.75\pm0.14 \\ 
            $\textbf{$\Modec$}$   & 8.69\pm0.13 & 7.74\pm0.12 & 7.82\pm0.12 & 7.83\pm0.12 & 7.85\pm0.12 & 7.94\pm0.12 & 7.89\pm0.12 \\ 
            $\textbf{$\Moded$}$   & 7.23\pm0.13 & 6.45\pm0.12 & 6.58\pm0.13 & 6.62\pm0.13 & 6.63\pm0.13 & 6.79\pm0.13 & 6.71\pm0.13 \\ 
            $\textbf{$\Modee$}$   & 8.25\pm0.12 & 8.06\pm0.12 & 8.07\pm0.12 & 8.08\pm0.12 & 8.09\pm0.12 & 8.10\pm0.12 & 8.10\pm0.12 \\ 
            $\textbf{$\Modeaa$}$  & 17.96\pm0.30 & 15.68\pm0.29 & 16.45\pm0.29 & 16.65\pm0.29 & 16.71\pm0.30 & 17.32\pm0.30 & 17.03\pm0.30 \\ 
            $\textbf{$\Modebb$}$  & 7.90\pm0.09 & 7.19\pm0.09 & 7.40\pm0.09 & 7.44\pm0.09 & 7.47\pm0.09 & 7.61\pm0.09 & 7.54\pm0.09 \\ 
            $\textbf{$\Modedd$}$  & 4.86\pm0.10 & 4.35\pm0.10 & 4.47\pm0.10 & 4.50\pm0.10 & 4.51\pm0.10 & 4.65\pm0.10 & 4.58\pm0.10 \\ 
            $\textbf{$\Modeaaa$}$ & 11.77\pm0.26 & 11.62\pm0.26 & 11.63\pm0.26 & 11.65\pm0.26 & 11.65\pm0.26 & 11.65\pm0.26 & 11.64\pm0.26 \\ 
            $\textbf{$\Modeccc$}$ & 9.76\pm0.24 & 8.74\pm0.23 & 8.96\pm0.23 & 9.02\pm0.23 & 9.04\pm0.23 & 9.26\pm0.23 & 9.15\pm0.23 \\ 
            $\textbf{$\Modeddd$}$ & 9.98\pm0.13 & 9.39\pm0.12 & 9.52\pm0.12 & 9.57\pm0.13 & 9.57\pm0.12 & 9.74\pm0.13 & 9.66\pm0.13 \\ 
            $\textbf{$\Modef$}$   & 21.31\pm0.54 & 21.05\pm0.54 & 21.08\pm0.54 & 21.08\pm0.54 & 21.11\pm0.54 & 21.14\pm0.54 & 21.14\pm0.54 \\ 
            \hline \hline
          \end{tabular}
          
          \label{tab:DTeff_xikspi}
	\end{center}
      \end{table*}  

The beam-constrained mass $M_{\rm{BC}}\equiv\sqrt{E_{\rm beam}^2/c^4-|{\vec{p}}|^2/c^2}$ is utilized to identify the $\lambdacm$ ST candidates, where $E_{\rm beam}=\sqrt{s}/2$ represents the average value of the $e^{+}$ and $e^{-}$ beam energies and $\vec{p}$ corresponds to the momentum of $\lambdacm$ candidates in the center-of-mass system of the $\ee$ collision. The $M_{\rm{BC}}$ of reconstructed $\lambdacm$ ST candidates is expected to peak around the $\lambdacm$ nominal mass which is approximately $2.286$~GeV.
 The energy difference $\Delta{}E \equiv E - E_{\rm beam}$ of the $\lambdacm$ ST candidates is used to improve the signal purity, where $E$ represents the total energy of the reconstructed $\lambdacm$ candidates. The corresponding requirement for the energy difference is in accordance with the criteria outlined in Ref.~\cite{BESIII:2022xne}.
In case of multiple $\lambdacm$ ST candidates in an event, the best $\lambdacm$ candidate is selected by choosing the one with the minimal $|\Delta E|$. The $M_{\rm BC}$ distribution for the 12~ST modes in the data sample is consistent with Ref.~\cite{BESIII:2022xne}.  Each mode has a distinct signal peak around the $\lambdacm$ nominal mass.
The unbinned maximum-likelihood fit is performed for the $M_{\rm BC}$ distribution to determine the ST yield, which aligns with the fit strategy reported in Ref.~\cite{BESIII:2022xne}.
Candidates within the $M_{\rm BC}$ signal region $(2.275,2.310)$~GeV/$c^{2}$ are retained for further analysis. The ST efficiencies are determined by analyzing the inclusive MC sample. The ST yields and detection efficiencies of the 12~modes at the seven energy points are presented in Tables~\ref{tab:STyields} and \ref{tab:STeff}, respectively.

\begin{figure}[t]
    \includegraphics[width=0.45\textwidth]{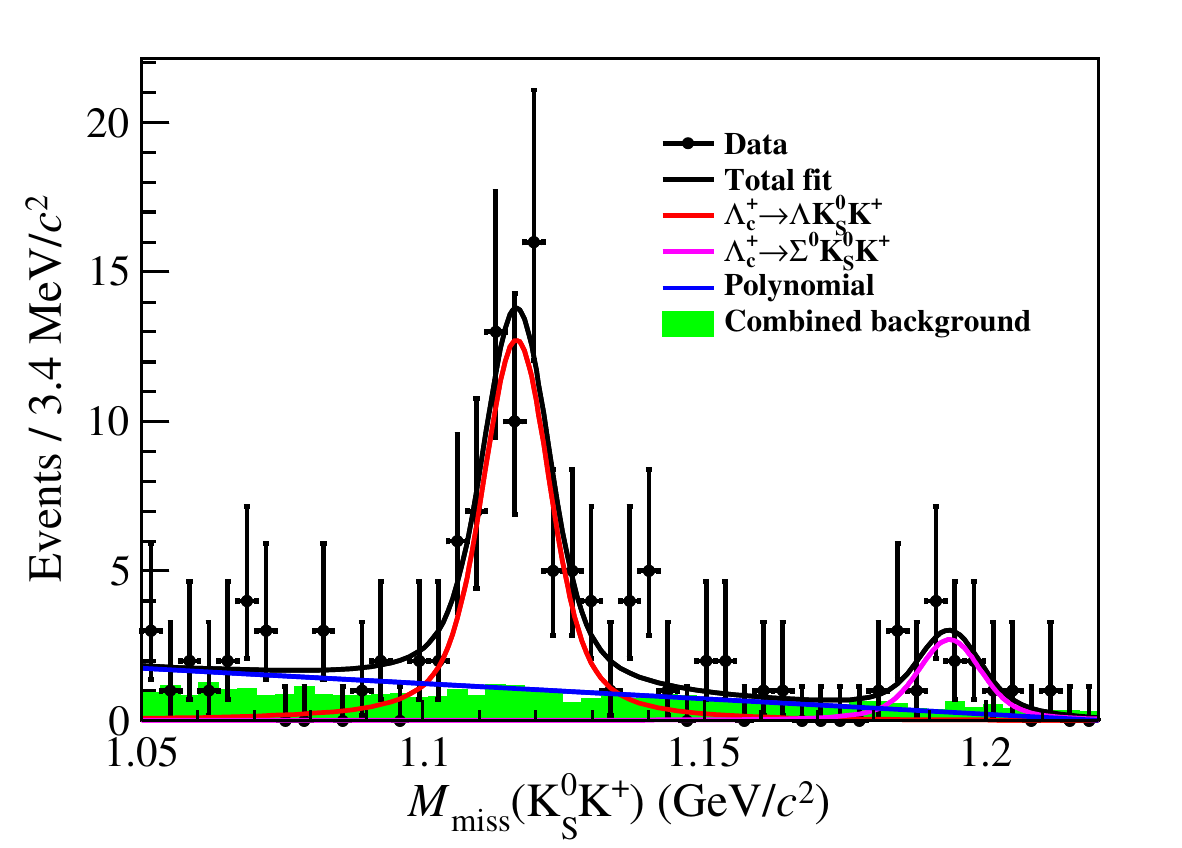}
    \caption{Fit result of the $M_{\rm miss}(K_{S}^{0}K^{+})$ distribution. The points with error bars are data. The solid black line is a sum of fit functions. The red and pink solid lines are the $\Lambda K_{S}^{0}K^{+}$ and $\Sigma^{0} K_{S}^{0}K^{+}$ signal shape, respectively. The blue solid line is the polynomial function and the green shaded histogram is the simulated background contribution derived from the inclusive MC sample.}
    \label{fig:fit_mm_ksk}
\end{figure}

\begin{figure}
    \centering
    \includegraphics[width=0.45\textwidth]{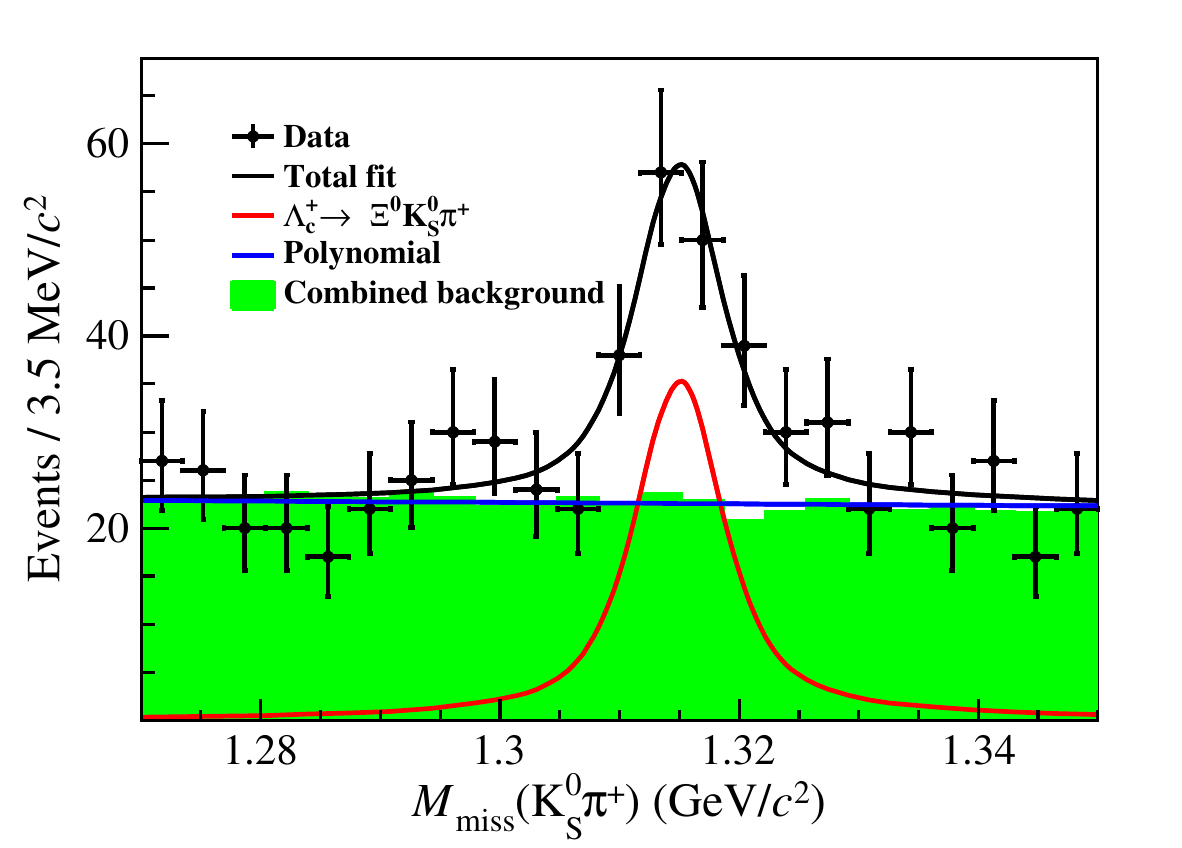}
    \caption{Fit result of the $M_{\rm miss}(K_{S}^{0}\pi^{+})$ distribution. The points with error bars are data. The black solid line is a sum of fit functions. The red solid line is the $\Xi^{0}K_{S}^{0}\pi^{+}$ signal shape. The blue solid line represents the polynomial function for background and the green shaded histogram is the simulated background contribution derived from the inclusive MC sample.}
    \label{fig:fit_mm_kspi}
\end{figure}

For the decays $\LctoLmdKsK$, $\LctoSigmaKsK$ and $\LctoXiKsPi$, the kinematic variable $M_{\rm miss} \equiv \sqrt{E_{\rm miss}^{2}/c^4-|\vec{{p}}_{\rm miss}|^{2}/c^2}$ is used to search for the $\Lambda$, $\Sigma^{0}$ and $\Xi^{0}$ candidates, respectively.
The $E_{\rm miss}$ and $\vec{{p}}_{\rm miss}$ are calculated by 
$E_{\rm miss} \equiv E_{\rm beam}-E_{\rm rec}$ 
and $\vec{{p}}_{\rm miss} \equiv \vec{{p}}_{\Lambda_{c}^{+}} - \vec{{p}}_{\rm rec}$, where $E_{\rm rec}$ and $\vec{{p}}_{\rm rec}$ are the energy and momentum of the reconstructed final state particles other than $\lambdacm$ in the $\ee$ center-of-mass system. The $E_{\rm beam}$ is used to replace the energy of the $\lambdacp$ candidate to improve the resolution. The $\Lambda_{c}^{+}$ baryon momentum $\vec{{p}}_{\Lambda_{c}^{+}}$ is calculated as $\vec{{p}}_{\Lambda_{c}^{+}} \equiv -\hat{p}_{\rm tag} \sqrt{E_{\rm beam}^{2}/c^2-m_{\Lambda_{c}^{+}}^{2} c^2}$, where $\hat{p}_{\rm tag}$ is the momentum direction of the $\bar{\Lambda}_{c}^{-}$ ST candidate and $m_{\Lambda_{c}^{+}}$ is the nominal mass of the $\Lambda_{c}^{+}$~\cite{ref:pdg2023}.
No significant peaking background contribution is observed in studies of the inclusive MC sample and the data sample within the $M_{\rm{BC}}$ and $M(K_{S}^{0})$ sideband region.
The $M_{\rm miss}$ distributions from the data sample, after combining all seven energy points and 12~tag modes, are depicted
in Figs.~\ref{fig:fit_mm_ksk} and \ref{fig:fit_mm_kspi}. In Fig.~\ref{fig:fit_mm_ksk} the two peaks around the nominal masses of the $\Lambda$ and $\Sigma^{0}$ represent the $\LctoLmdKsK$ and $\LctoSigmaKsK$ decays, respectively, while Fig.~\ref{fig:fit_mm_kspi} shows the $\LctoXiKsPi$ decay. The $\Lambda$, $\Sigma^{0}$ and $\Xi^{0}$ signal shapes are modeled by individual MC-derived shapes convolved with a Gaussian function with free parameters. Since no peaking background contribution is observed,
a linear function is used to describe the total background contributions, which  includes $q\bar{q}$ events from continuum hadron production in $\ee$ annihilation and $\lambdacp\lambdacm$ pairs from $\ee\to\lambdacp\lambdacm$ events, excluding the corresponding signal processes.
The fitted signal yields for the decays $\LctoLmdKsK$, $\LctoSigmaKsK$ and $\LctoXiKsPi$ are $67.7\pm 10.1$, $13.1^{+4.6}_{-3.9}$ and $125.4\pm 20.2$, respectively, where the uncertainty is statistical. The statistical significances are $9.4\sigma$, $3.4\sigma$ and $6.6\sigma$ for the signal modes $\LmdKsK$, $\SigmaKsK$, and $\XiKsPi$, respectively, and are estimated by comparing the fit likelihoods with and without including the signal components.
\begin{figure}[t]
    \includegraphics[width=0.45\textwidth]{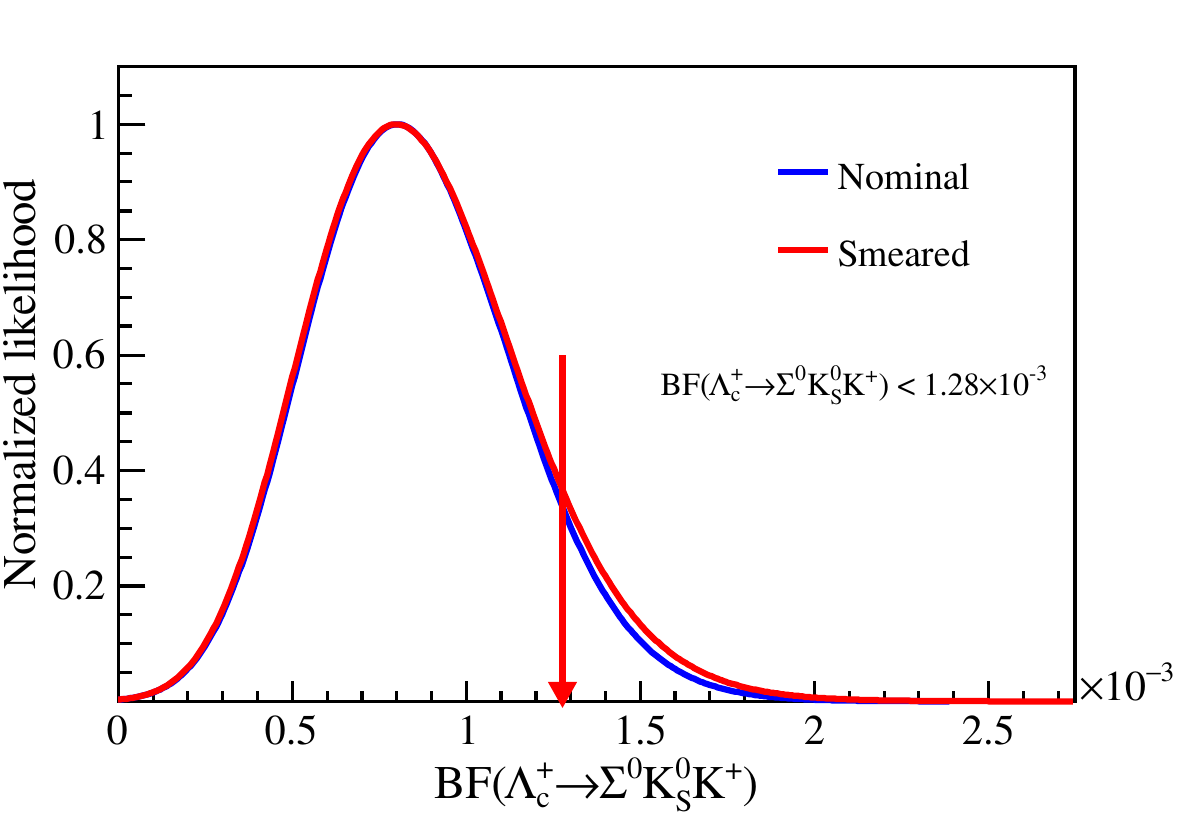}
    \caption{Normalized likelihood distribution versus the BF of $\LctoSigmaKsK$. The blue solid line is the nominal distribution. The red solid line is the smeared distribution with systematic uncertainties considered. The red arrow represents the final upper limit result.}
    \label{fig:upper_limit}
\end{figure} 
The BFs of the signal decays are calculated as: 
\begin{equation}
\label{eq:br}
\mathcal{B}=\frac{N^{\rm DT}}{\sum_{ij} N_{ij}^{\mathrm{ST}}\cdot (\epsilon_{ij}^{\mathrm{DT}}/\epsilon_{ij}^{\mathrm{ST}})\cdot\mathcal{B}_{\rm{int}} },  
\end{equation}
where $i$ and $j$ are the different ST modes and center-of-mass energies of the data samples. The $N^{\rm DT}$ is the total DT yield of each ST mode and center-of-mass energy. The $\mathcal{B}_{\rm{int}}$ is the BF of $K_{S}^{0}\to\pi^{+}\pi^{-}$, which is $(69.2\pm0.05)\%$ as quoted by the PDG~\cite{ref:pdg2023}. The $N_{ij}^{\mathrm{ST}}$, $\epsilon_{i,j}^{\mathrm{ST}}$, and $\epsilon_{i,j}^{\mathrm{DT}}$ represent the ST yield and efficiency, and DT efficiency, respectively, where the latter is obtained by using the corresponding exclusive DT signal MC samples.
 The efficiencies $\epsilon_{i,j}^{\mathrm{ST}}$, $\epsilon_{i,j}^{\mathrm{DT}}$ for the 12~tag modes for each center-of-mass energy are presented in Tables~\ref{tab:STeff}-\ref{tab:DTeff_xikspi}. The corresponding BF results are summarized in Table~\ref{tab:Th_bf_vs_exp}. Since the $\LctoSigmaKsK$ decay has a relatively low statistical significance, the upper limit on the BF is set at the $90\%$ confidence level based on the Bayesian approach with a flat prior~\cite{Feldman:1997qc, Stenson:2006gwf, ref:upper, BESIII:2021drk}.  The upper limit is determined to be $1.28\times10^{-3}$ as shown in Fig.~\ref{fig:upper_limit}.
 Corresponding systematic uncertainties are taken into account in the estimation of the upper limit. The additive uncertainties, represented by the $M_\text{miss}$ fit in the Table~\ref{tab:sys_err}, are addressed by varying the DT fit method and the most conservative upper limit is chosen.
 The multiplicative uncertainties are used to smear the likelihood distribution by a value of $8.6\%$.

\section{systematic uncertainties}
\begin{table*}[t]
  \begin{center}
    \caption{Relative systematic uncertainties (in $\%$) for the BF measurements. }
  \begin{tabular}{l|c|c|c}
      \hline \hline
			Source            &  $\LctoLmdKsK$ &  $\LctoSigmaKsK$ & $\LctoXiKsPi$  \\
			\hline
			Tracking          & 1.0 & 1.0 & 1.0 \\
			PID               & 1.0 & 1.0 & 1.0  \\
			$M_{\rm miss}$ fit           & 2.1 & 17.6 & 0.8  \\
            $K_{S}^{0}$ reconstruction   & 2.6 & 3.8 & 1.6 \\
            $\mathcal{B}(K_{S}^{0}\to\pi^{+}\pi^{-})$ & 0.1 & 0.1 & 0.1 \\
			$\lambdacm$ ST yields & 0.7 & 1.2 & 1.2 \\
			MC model          & 2.9 & 7.5 & 5.1 \\
            MC statistics     & 0.3 &  0.3  &  0.3 \\
			\hline
      Total             &4.7 &19.6 & 5.7     \\
      \hline \hline
    \end{tabular}
    \label{tab:sys_err}
  \end{center}
  \end{table*}

The sources of systematic uncertainty for the BF measurements include track and $K_{S}^{0}$ reconstruction, PID, the $M_{\rm miss}$ fit, the quoted intermediate BF of the $K_{S}^{0}$, determination of the $\lambdacm$ ST yields, choice of the MC model, and generated MC statistics.

The systematic uncertainties from tracking reconstruction and PID of the charged kaon and pion are $1\%$, as quoted from Ref.~\cite{BESIII:2022xne}. The systematic uncertainty due to the fitted DT yields is estimated by altering the shape of the signal and background candidates in the fit of the $M_{\rm miss}$ distribution. For the description of the signal shape, the MC derived curve is convolved with a double Gaussian function while the background shape is changed from a 1st-order polynomial function to a MC derived shape. For the $K_{S}^{0}$ reconstruction, the systematic uncertainty due to the efficiency difference between MC and data samples is estimated using a control sample of the decays $J/\psi\to K_{S}^{0}K^{\pm}\pi^{\mp}$. The efficiency of $K_{S}^{0}$ reconstruction is recalculated based on
 \begin{equation}
     \label{eq:Ksrec1}
   \varepsilon'=\frac{1}{N}\sum_{i=1}^{n}\left(\frac{\varepsilon^{\mathrm{data}}(P_{i})}{\varepsilon^{\mathrm{MC}}(P_{i})}\right)
 \end{equation}
 where $\varepsilon'$ is the reweighted detection efficiency of the $K_{S}^{0}$ obtained from a control sample of the decays $J/\psi \to K_{S}^{0} K^{\pm} \pi^{\mp}$ of data or simulated samples, respectively, depending on the $K_{S}^{0}$ momentum, $P$.
 The $N$ is the total number of events in the MC sample while $n$ is the selected number of signal events. The uncertainty is determined by comparing the nominal BF result and the BF with recalculated efficiency. The BF of $K_{S}^{0}\to\pi^{+}\pi^{-}$ is $(69.20\pm 0.05)\%$ quoted from the PDG~\cite{ref:pdg2023}, which contributes to a relative systematic uncertainty of $0.1\%$. The uncertainty from the $\lambdacm$ ST yield is estimated by  varying the mass range and the signal and background shapes in the fit of the $M_{\mathrm{BC}}$ distribution. To estimate the uncertainty from the MC model, the distributions of the simulated sample are reweighted to agree with the data distributions. The mass distribution of $K_{S}^{0}K^{+}$, $\Lambda K^{+}$ and $\Lambda K_{S}^{0}$ for the decay $\LctoLmdKsK$  and the mass distributions of $K_{S}^{0}K^{+}$, $\Sigma^{0}K^{+}$ and $\Sigma^{0}K_{S}^{0}$ for the decay $\LctoSigmaKsK$  obtained from the $M_{\mathrm{BC}}$ sideband subtraction method are used to reweight the PHSP signal MC distributions. In case of the decay $\LctoXiKsPi$, the mass distribution of $K_{S}^{0}\pi^{+}$, $\Xi^{0}\pi^{+}$ and $\Xi^{0}K_{S}^{0}$ obtained using the $S$-weights method~\cite{Pivk:2004ty} is utilized to reweight the PHSP signal MC distributions. The difference of BF between the nominal and reweighted samples is assigned as a systematic uncertainty.
The uncertainty from the statistical fluctuation $\Delta_{\mathrm{sys}}^{\mathrm{MC}}$ of the MC sample is estimated to be $0.3\%$ for the signal decay modes according to:
\begin{equation}
\label{eq:4}
        \Delta_{\mathrm{sys}}^{\mathrm{MC}} = \frac{\sqrt{\sum_{j}(N_{j}^{\mathrm{ST}}\Delta\varepsilon^{j})^{2}}}{\sum_{j}N_{j}^{\mathrm{ST}}\varepsilon^{j}}, 
    \end{equation} 
where $j$ is the energy point, $N_{j}^{\mathrm{ST}}$ is a sum of ST yields from all tag modes at each energy point $j$, $\epsilon^{j}=\sum_{i}(\frac{N_{ij}^{\mathrm{ST}}}{\epsilon_{i,j}^{\mathrm{ST}}}\cdot\epsilon_{i,j}^{\mathrm{DT}})/\sum_{i}N_{ij}^{\mathrm{ST}}$ is the reduction efficiency and $\Delta\epsilon^{j}$ is the corresponding error.

Assuming that all the sources are uncorrelated, the total systematic uncertainties of the three decay modes are obtained by adding each of the sources in quadrature as listed in Table~\ref{tab:sys_err}.

\section{summary}

The three Cabibbo-favored decays $\LctoLmdKsK$, $\LctoSigmaKsK$ and $\LctoXiKsPi$ are studied by analyzing a 4.5~fb$^{-1}$ data sample at seven center-of-mass energy points varying from 4599.53~MeV to 4698.82~MeV. The BFs of the decays $\LctoLmdKsK$, $\LctoSigmaKsK$ and $\LctoXiKsPi$ are measured to be $(3.12\pm0.46\pm0.15)\times10^{-3}$, $(0.80^{+0.28}_{-0.24}\pm0.16)\times10^{-3}$ and $(3.70\pm0.60\pm0.21)\times10^{-3}$, respectively, where the first uncertainty is statistical and the second is systematic.
The BF upper limit of the decay $\LctoSigmaKsK$ is determined for the first time and is set to be $1.28\times10^{-3}$ at the $90\%$ confidence level. The measured BF of the decay $\LctoLmdKsK$ is consistent with the PDG value $(2.80\pm0.55)\times 10^{-3}$ with improved precision. This result is also consistent with the previous BESIII measurement~\cite{ref:27} and the theoretical predictions~\cite{Geng:2018upx, Cen:2019ims}. The combination with the previous BESIII measurement~\cite{ref:27} gives the average BF of the decay $\LctoLmdKsK$, $(3.07\pm0.26\pm0.13)\times10^{-3}$, taking into account the small overlap between the signal samples in the two analyses.

The decay $\LctoXiKsPi$ is observed for the first time and the measured BF is about one order of magnitude lower than the theoretical predictions~\cite{Geng:2018upx, Cen:2019ims}, but is consistent with the statistical isospin model calculation~\cite{ref1} and the theoretical predicition incorporating the contribution of the $H(15)$~\cite{Geng:2024h5}. The BF upper limit of the decay $\LctoSigmaKsK$ is consistent with Ref.~\cite{Cen:2019ims}, but it is incompatible with Ref.~\cite{Geng:2018upx}. These discrepancies motivate further investigations to enhance our understanding of the $\Lambda_{c}^{+}$ decays involving more than one strange hadron in the final state. All the results are dominated by the statistical uncertainty. Larger datasets are planned to be collected near the $\Lambda_{c}^{+}\bar{\Lambda}_{c}^{-}$ threshold in the upcoming years~\cite{ref:28} and will allow for further investigation and will deepen our knowledge of the decay mechanisms of charmed baryons.

\acknowledgments
The BESIII Collaboration thanks the staff of BEPCII and the IHEP computing center for their strong support. This work is supported in part by National Key R\&D Program of China under Contracts Nos. 2020YFA0406400, 2020YFA0406300, 2023YFA1609400, 2023YFA1606000; National Natural Science Foundation of China (NSFC) under Contracts Nos. 12105127,  12422504, 11635010, 11735014, 11935015, 11935016, 11935018, 12025502, 12035009, 12035013, 12061131003, 12192260, 12192261, 12192262, 12192263, 12192264, 12192265, 12221005, 12225509, 12235017, 12361141819; the Chinese Academy of Sciences (CAS) Large-Scale Scientific Facility Program; the CAS Center for Excellence in Particle Physics (CCEPP); Joint Large-Scale Scientific Facility Funds of the NSFC and CAS under Contract No. U1832207; 100 Talents Program of CAS; Fundamental Research Funds for the Central Universities, Lanzhou University; The Institute of Nuclear and Particle Physics (INPAC) and Shanghai Key Laboratory for Particle Physics and Cosmology; Agencia Nacional de Investigación y Desarrollo de Chile (ANID), Chile under Contract No. ANID PIA/APOYO AFB230003; German Research Foundation DFG under Contract No. FOR5327; Istituto Nazionale di Fisica Nucleare, Italy; Knut and Alice Wallenberg Foundation under Contracts Nos. 2021.0174, 2021.0299; Ministry of Development of Turkey under Contract No. DPT2006K-120470; National Research Foundation of Korea under Contract No. NRF-2022R1A2C1092335; National Science and Technology fund of Mongolia; National Science Research and Innovation Fund (NSRF) via the Program Management Unit for Human Resources \& Institutional Development, Research and Innovation of Thailand under Contracts Nos. B16F640076, B50G670107; Polish National Science Centre under Contract No. 2019/35/O/ST2/02907; Swedish Research Council under Contract No. 2019.04595; The Swedish Foundation for International Cooperation in Research and Higher Education under Contract No. CH2018-7756; U. S. Department of Energy under Contract No. DE-FG02-05ER41374.

\end{document}